\documentstyle[prl,twocolumn,aps,psfig]{revtex}

\def\bvec#1{{\rm\bf #1}}

\def\mf{\langle u_t\rangle}
\def\mi{\langle u_i\rangle}
\def\bea{\begin{eqnarray}}
\def\eea{\end{eqnarray}}
\def\be{\begin{equation}}
\def\ee{\end{equation}}
\def\eps{\epsilon}
\def\dst{\displaystyle\phantom{|}}
\def\ov{\over\displaystyle}

\begin{document}
\title{Simple analytic solution of fireball hydrodynamics}
\author{
T. Cs{\"o}rg{\H o}\\
MTA KFKI RMKI, H-1525 Budapest 114, POB 49, Hungary}
\maketitle
\begin{abstract}
A new family of simple analytic solutions of hydrodynamics is found 
for slowly expanding, rotationally symmetric fireballs
assuming an ideal gas equation of state.
The temperature profile is position
independent only in the  collisionless gas limit.
The Zim\'anyi-Bondorf-Garpman solution and
the Buda-Lund parameterization of expanding hydrodynamic
particle sources  are recovered as special cases.
The results are applied to predict
new features of proton correlations and
spectra at 1.93 AGeV $Ni+Ni$ collisions.
\end{abstract}

{\it Introduction} ---
Hydrodynamical models are applied in a wide range of physical
problems from the evolution of galactic systems through the explosion
of supernovas to the description of particle spectra and correlations
in heavy ion collisions in a broad energy range. Due to the nonlinearity
of the hydrodynamical equations it is rather difficult
to solve them in an analytic manner.
In particle and  heavy ion physics some observables
like the single-particle spectra and the two-particle Bose-Einstein
correlation functions are sensitive to the means and variances of
the hydrodynamical flow and density profiles~\cite{3d,xuprof}.
This result implies that a hydrodynamical solution involving only
the means and the variances of the density profile,
flow profile and temperature profile is of great
interest. Such a solution of the equations
of  non-relativistic (NR) hydrodynamics is presented herewith.
Physical realization of the solution is possible both by
heavy ion collisions in the 30 MeV A - 2 GeV A energy domain~\cite{nr,nrt,FOPI}
and in a more limited sense by the NR transversal flows
that seem to develop in high energy hadron-hadron and nucleus-nucleus
collisions~\cite{3d,na22,na44sp}.

{\it The new family of exact solutions} --- 
Let us consider the NR hydrodynamical problem,
as specified by the continuity, Euler and energy equations:
\bea
{\partial_t n} + \bvec\nabla (\bvec v n) & = & 0, \label{e:cont}\\
{\partial_t \bvec v} + (\bvec v\bvec\nabla) \bvec v &  = &
		- (\bvec\nabla p) / (m n) \label{e:Eu},\\
{\partial_t \eps } + \bvec\nabla (\eps \bvec v )
		& = & - p \bvec\nabla \bvec v ,
		\label{e:en}
\eea
where $n$ denotes the number density of particles,
$\bvec v$ stands for the NR flow field,
$\eps $ for the NR energy density,
$p$ for the pressure and in the following the temperature
field is denoted by $T$.
We assume a NR ideal gas equation of state,
\bea
	p & = & n T, \label{e:p}\\
	\eps & = & {3 \ov 2} p \, = \, {3 \ov 2 } n T,\label{e:eos}
\eea
which closes the set of equations for $n$, $\bvec v$ and $T$.

We are looking for a set of solutions for a self-similar
density, flow  and temperature profile. Our ansatz for the
self-similarity means that the non-trivial dependence of the
solution on the spatial and temporal variables happens
through the {\it scaling variable}
\be
x = \bvec r^2 / R^2(t),
\ee
where $R(t)$ is a characteristic, time dependent radial scale.
Note that the scaling variable $x$ is not to be confused with $r_x$,
the first component of the space vector $\bvec r = (r_x, r_y, r_z)$.
Let us assume rotational symmetry and
introduce the dimensionless scaling functions $\nu(x)$,
$\beta(x)$ and ${\cal T}(x)$ instead of $n = n(\bvec r, t)$,
$\bvec v = \bvec v(\bvec r,t)$ and $T = T(\bvec r, t)$, respectively:
\bea
	n & = & n_0\, {R_0^3 \ov R^3} \, \nu(x), \label{e:n} \\
	\bvec v & = & f_v(t) \, \beta(x) \,\bvec r,  \\
	T & = & T_0 \, f_T(t) \, {\cal T}(x) ,
\eea
	the time-dependence of the radial scale 
	is written as
\bea
	R^2 & = & R^2(t) \, = \, R_0^2 \, \phi(t), \label{e:rphi}
\eea
	and $n_0$, $T_0$ and $R_0$  are a constants
	related to the initial conditions.
	The form of this  ansatz is not modified if we specify
	that the minimal ``turning point radius" $R_0$ is prescribed
	 at a given instant $t_0$  and the central particle 
	density and temperature  at this  time $t_0$ is 
	$n_0 = n(t_0, \bvec 0)$ and $T_0 = {T}(t_0, \bvec 0)$.
	This choice of constants implies 
\bea
	\nu(0) & = & \beta(0) \, = \, {\cal T}(0) \, = \, 1, \\
	f_T(t_0) & =  & \phi(t_0) = 1, \qquad R_0 \, = \, R(t_0).
\eea
	The unknown functions $f_v(t)$, $f_T(t)$, $\phi(t)$  
	are determined as follows.
	{}From the continuity equation one obtains that
\be
	f_v(t) \, = \, {\dst R^{\prime} \ov R},
	\qquad \mbox{\rm and} \quad \beta(x) = 1,
\ee
	where ordinary differentiation according to the argument
	is denoted by $^{\prime}$.
	Thus the continuity equation implies
\be
	\bvec v = \bvec r {\dst R^{\prime}\ov R},\label{e:hh}
\ee
	regardless of the shape of the density profile.

	Utilizing the equation of state and the continuity
	equation, the energy equation
	can be rewritten as
\be
	\partial_t T + (\bvec v \bvec \nabla) T +
		{2\ov 3} T \bvec \nabla \bvec v \ = \ 0.
		\label{e:ener}
\ee
	Inserting the ansatz into this form and making use of eq.~(\ref{e:hh})
	we obtain the time-dependence of the $T$ as 
\be
	f_T(t) \, = \, {R_0^2 \ov R^2} \, = \, {1 \ov \phi(t)}.
		\label{e:ttsol}
\ee
	Inserting this to the Euler equation, we obtain
\bea
	\phi^{\prime\prime} \phi - 0.5 (\phi^{\prime})^2 & = &
		{2 T_0 \ov m  R_0^2} C_{\phi}, \label{e:fifi} \\
	{\cal T}^{\prime} + {\cal T} {\nu^{\prime}\ov \nu} & = &
		- { C_{\phi} \ov 2}. \label{e:nn}
\eea
	It is trivial to solve these equations for $\nu(x)$ and $\phi(t)$.
	The solution in terms of physical parameters reads as
\bea
	\phi(t) & = & 1 + {\mf^2 \ov R_0^2} (t - t_0)^2 ,
	\qquad C_{\phi} \, = \, {\dst m\ov T_0} \mf^2, 
			\label{e:fisol}\\
	\nu(x) & = & { 1  \ov {\cal T}(x) }
	\exp\left( -  {\dst m \mf^2 \ov 2 T_0 }
		\int_0^x {\dst du \ov {\cal T}(u)} \right).
			\label{e:hsol}
\eea
	The mean radial flow parameter $\mf$
	controls the {\it asymptotic} speed of the expansion,
	as $\lim_{t \rightarrow \infty} R^{\prime}(t) = \mf$ .
	In the above solutions, the functional form of ${\cal T}(x)$, the
	scaling function of the radial temperature profile,
	can be chosen freely, corresponding to the freedom in the
	choice of the initial boundary conditions.
	These new hydro solutions read  in terms of physical variables
	as :
\bea
	R^2(t) & = & R^2_0 + \mf^2 (t - t_0)^2, \label{e:rtsol} \\
	\bvec v(t,\bvec r) & = &
		 {R^{\prime}(t)\ov R(t)} \, \bvec r \, = \,
		{\mf^2 (t - t_0) \ov R^2_0 + \mf^2 (t-t_0)^2}\, \bvec r ,
		\label{e:vsol} \\
	T(t,\bvec r) & = & T_0\, {\cal T}(x) \, {R_0^2 \ov R^2} \, = \,
	T\left[t_0, \bvec r R_0 / R(t)\right]\,
		{R_0^2 \ov R^2(t)} \label{e:tsol} \\
	n(t,\bvec r) & = & n_0 \, {\dst R_0^3 \ov R^3(t)} \,
		{\dst T(t_0,\bvec 0) \ov
		T\left[t_0, \bvec r  R_0 / R(t)  \right]}
		\, \times \nonumber \\
	\null &  \times &
			\exp\left[
		- m \mf^2 {\dst R^2(t) \ov R_0^2}
		\int_0^{|\bvec r|}
		r dr {\dst T(t_0, 0) \ov T(t_0, r) }
			\right]
		. \label{e:nsol}
\eea
	This solution describes a
	system that contracts for $t < t_0$, reaches its
	minimum size $R_0$ at the time of the turning point $t = t_0$
	when the flow field vanishes. For later times, $t > t_0$,
	the system expands again.
	The temperature field is obtained by rescaling the initial
	temperature field.

	The above explicit solution for the density profile can
	be written into a simpler form using the scaling
	variable $x$ and the scaling function of
	the temperature profile, ${\cal T}(x)$:
\be
	n(t,\bvec r) \, = \, { n_0  \ov {\cal T}(x) }
		\, {\dst R_0^3 \ov R^3(t)}\,
		\exp\left[ - {\dst  m \mf^2  \ov 2 T_0}
		\int_0^x {\dst du \ov {\cal T}(u) }
		\right].
		\label{e:nsimpl}
\ee
	This completes the generic description of our solution
	of NR spherically symmetric fireball
	hydrodynamics. The solution, given by eqs.~
	(\ref{e:rtsol}-\ref{e:nsimpl}), depends on
	4 constants of integration, $R_0$, $\mf$, $T_0$ and
	$t_0$, as well as on the dimensionless {\it function }
	${\cal T}(x)$.  Three of the constants of integration
	correspond to the three integrated equations 
	(continuity, Euler and energy equation). The
	constant $t_0$ corresponds to the homogeneity
	of the equations of NR hydrodynamics
	in time. The free choice of the dimensionless
	scaling function corresponds to the freedom in
	the specification of the 
	initial boundary conditions.
	Thus we have found infinitely many new solutions
	of NR hydrodynamics, one for each integrable function
	of one variable. Note that each of these solutions 
	is scale invariant: 
\bea
	n\left(t, \bvec r {R \ov R_0} \right) \,
		{ R^3 \ov R^3_0}
		  &  = & n(t_0, \bvec r), \label{e:nscale}  \\
	\bvec v\left(t, \bvec r {R \ov R_0} \right)
		{R^{\prime}_0 \ov R^{\prime} } \,
		  &  = & \bvec v (t_0, \bvec r),  \label{e:vscale} \\
	T\left(t, \bvec r {R \ov R_0} \right) \,
		{ R^2 \ov R^2_0}
		  &  = & T(t_0, \bvec r) , \label{e:tscale}
	\label{e:sinv}
\eea
	independently of the time $t$, where $R^{\prime}_0 = R^{\prime}(t_0)$. 

	The self-similarity can be utilized to match this solution
	to any given initial time $t_i$ for a given initial radius $R_i$
	and a given (rotationally symmetric) temperature
	distribution $T_i(\bvec r)$ at time $t_i$.
	These initial conditions are
\bea
	T_i(\bvec r)
		&  = & 
	T(t_i,\bvec r), \quad
	R(t_i)\, =\, R_i, \\ 
	\bvec v(\bvec r,\,t_i) & = & {\bvec r\ov R_i} \mi,
	\quad n(t_i,\bvec 0) \, = \, n_i.
\eea
	This yields the following solutions for $\phi$ and $R(t)$:
\begin{eqnarray}
	\phi_i(t) & = & \left(1 + {\mi\ov R_i}(t-t_i)\right)^2
		+ {\mf^2\ov R_i^2} (t-t_i)^2, \label{e:fisol2}\\
	R(t)^2 & = & R_i^2  \phi_i(t), \qquad T_i \, = \, T_i(\bvec 0).
\end{eqnarray}
	This solution appears instead of eqs. ~(\ref{e:fisol}, \ref{e:rtsol})
	if the initial temperature field, initial radius and initial flow field
	is prescribed instead of the turning point temperature and radius.

	The new solution presented above also solves the
	equation of state of an ideal gas for isentropic expansion~\cite{jnr}:
\be
	\left[\partial_t +(\bvec v \bvec\nabla) \right] p n^{-5/3} \,  = \, 0.
\ee
	Note that {\it the time-dependence of the radius and
	the solution for the flow-field is independent of the initial
	temperature profile} $T(t_0, \bf r)$.
	Due to this property of our general solution ,
	we can study various initial conditions in a simple manner,
	assuming various distributions of $T(t_0, \bf r)$
	and evaluate its influence on the density profile only.

{\it Knudsen-gas limit} ---
	An essential feature of our solution is  that the
	central and the asymptotic temperatures may differ and induce
	a radial dependence of the temperature field. If the
	initial temperature field is constant in space,
	we find a special case of ${\cal T}(x) = 1$ and
	$T(t_0, \bvec r) = T(t_0) = T_0$.
	Eq.~(\ref{e:nsimpl}) yields that the density profile is Gaussian
	with $R_G^2  =  R^2 T_0 / (m \mf^2)$:
\be
	n = {\dst N \ov [2 \pi R_G^2(t)]^{3/2} }
		\exp\left( - {\bvec r^2 \ov 2 R_G^2(t)} \right).
\ee
	It is straightforward to prove ~\cite{cspeter} by evaluating
	the time dependence of the phase-space density,
	that this case corresponds to a collisionless, 
	free streaming of the particles. Hence this parameterization can
	be applied only to describe the post-freeze-out stage of
	heavy ion collisions.
	Thus, {\it non-trivial hydrodynamical evolution
	takes place  in the presence of a non-vanishing
	temperature gradient only},
	for  all rotationally symmetric solutions of the non-relativistic
	hydrodynamical equations that we found.

{\it The Zim\'anyi-Bondorf-Garpman solution} ---
	20 years ago,
	a famous analytic solution of NR hydrodynamics
	was found by Zim\'anyi, Bondorf and Garpman (ZBG) in ref.~\cite{jnr}.
	The temperature and the density for that
	 solution explicitly vanishes if $|\bvec r|$ is
	larger than an expanding radial scale, $ |\bvec r| > R(t)$.
	The ZBG solution has been extensively applied
	to describe particle spectra for NR heavy ion collisions.
	The ZBG solution has the following simple initial temperature
	profile:
\be
	T(t_0, \bvec r) = T_0
		\left[ 1 - {\dst \bvec r^2 \ov R^2(t)}\right]
	\qquad \mbox{\rm if}\,\,\, |\bvec r| \le R(t).
\ee
	Inserting this to our general solution,
	eq.~(\ref{e:nsimpl}), we obtain
\be
	n(t,\bvec r) = n_0
		{\dst R_0^3 \ov R^3}
		\left[ 1 - {\dst \bvec r^2 \ov R^2}\right]^{C_\phi /2 -1}
	\quad \mbox{\rm if}\,\,\, |\bvec r| \le R(t),
\ee
	which corresponds to the ZBG solution with  an exponent
	$\alpha = C_\phi/2 -1 = {\dst m \mf^2  \ov 2 T_0} - 1$.

{\it Generalization to $d$-dimensional expansion} --
	Our analytic solution can be generalized to expansions in
	$d$ dimensions, where $d$ is an arbitrary number.
	The case $d = 2$ is especially interesting as it corresponds
	to the time evolution  of hot and dense hadronic
	matter in the transverse directions in high energy
	heavy ion~\cite{3d,qm95,s96} and in particle reactions~\cite{na22}.
	When analyzing particle correlations and spectra in these reactions,
	it is found that the longitudinal expansion is relativistic,
	(sometimes extremely relativistic), however, this longitudinal
	flow can be approximately decoupled from the transversal dynamics,
	which is found to be non-relativistic~\cite{s96,3d}.

	In $d$ dimensions, the divergences of the flow and the equation
	of state are modified as
\be
	\bvec \nabla \bvec r_d \,  = \, d, \qquad
	\eps_d  \, = \, {\dst d\over 2} p.
\ee
	The scaling variable is $x = \bvec r_d^2 / R^2(t)$,
 	the ansatz is	
\be
	n \, = \, n_0 \, {R_0^d \ov R^d(t)} \nu(x) \quad
	\bvec v_d \, = \, \bvec r_d f_d(t) \beta(x).
\ee
	The equations for the temperature and the radial scale parameter
	are unchanged.  In $d$ dimensions, the energy equation eq.~(\ref{e:ener})
	is modified to
\be
	\partial_t T + (\bvec v_d \bvec \nabla) T +
		{2\ov d} T \bvec \nabla \bvec v_d \ = \ 0.
		\label{e:tt}
\ee
	From this and the continuity equation we obtain after
	eliminating the singular solutions that
\be
	\bvec v_d = \bvec r_d {\dst R^{\prime}(t) \over R(t)}
\ee
	and the dimension of the expansion, $d$, cancels from the Euler
	equation. The solutions for
	$d$ dimensional expansions are very similar to the $d = 3$ dimensional
	expanding fireballs, the differences are that in
	eqs.~(\ref{e:vsol},\ref{e:tsol})
	one has to utilize $d$ dimensional vectors 
	for coordinate-space and flow,
	and in eq.~(\ref{e:nsol}) the density decreases as 
	$ n \propto R(t)^{-d}$.

{\it The Buda-Lund parameterization} ---
	The Buda-Lund parameterization (BL) was introduced to
	describe phenomenologically the NR 
	transversal flows appearing in high energy hadron-hadron
	and heavy ion reactions, given first in refs.~\cite{3d,qm95}.
	This parameterization keeps only the means and the variances
	of the density distribution, the inverse temperature profile
	and it corresponds to a scaling transversal flow, such as 
	eq.~(\ref{e:vsol}). 
	In refs.~\cite{3d,nrt,na22,s96}, a radial 
	temperature profile was used,that can be rewritten as:
\be
	T(\tau, {\bvec  r}) \! = \! {\dst T_0  
		\ov \left[{ 1 \! + \!
		\langle{\Delta T\ov T}\rangle_r {\bvec r^2\ov 2 R^2}}
		\right]
		\left[1 \! + \! \langle {\Delta T\over T} \rangle_\tau
		{(\tau - \tau_0)^2 \ov 2\Delta\tau^2}\right]},
		\label{e:tbl} 
\ee
	where $\bvec r = (r_x, r_y)$, the change of the temperature
	in the radial and the temporal directions is controlled by
	the free parameters $\langle \Delta T/T\rangle_r$
	and $\langle \Delta T/T\rangle_\tau$, respectively,
	and $C$ is a constant of normalization.
	In the above parameterization,
	the longitudinal proper-time, $\tau = \sqrt{t^2 - r_z^2}$ is
	introduced instead of the time $t$ ,
	corresponding to a relativistic longitudinal expansion. 
	The exact Buda-Lund density profile is obtained from 
	eqs.~(\ref{e:nsimpl},\ref{e:tbl}) as
\bea
	n(t_0,\bvec r) & = & n_0 \,
		\left({ 1 +
		\langle \Delta T/ T \rangle\phantom{|}_r 
		{\bvec r^2/ (2 R_0^2)}}
		\right) \, \times \nonumber \\
\null & \null &
		\! \exp\left[\! - {\dst m \mf^2 \ov 2 T_0}  \!
			\left( {\dst \bvec r^2 \ov  R_0^2} 
		+ \langle {\dst \Delta T\ov T} \rangle\phantom{|}_r  \,
		{\dst \bvec r^4 \ov R_G^4} \right) \right].
		\label{e:nbl}
\eea
	The time-evolution of this density can be obtained
	from simple re-scaling, eq.~(\ref{e:nscale}),
	and comparing 
	the BL ansatz for the time dependence of the 
	cooling, eq.~(\ref{e:tbl})
	 with the exact hydro solution we find
\be
	{\dst 1 \ov 2 \Delta \tau^2} \langle \Delta T/T\rangle_\tau
	\, = \, {\dst \mf^2 \ov R_0^2}. \label{e:cool}
\ee
	The exact density distribution in eq.~(\ref{e:nbl})
	can be approximated by a Gaussian, corresponding to
	the BL Gaussian ansatz for the density profile in
	ref.~\cite{3d}, if terms of
	${\cal O}(\bvec r^4/R^4)$ can be neglected.
	(Note that the difference between the BL and the ZBG solutions 
	is only ${\cal O}(\bvec r^4/R^4)$ if one sets
	$\langle\Delta T/T\rangle_r = 1$ in the BL hydro solution.)
	The BL Gaussian radius is 
\be
	{\dst 1 \ov R_G^2 }  \, = \,  {\dst 1 \ov R_0^2} \,
		\left( {\dst m \mf^2 \ov T_0 } -
		\langle \Delta T/T\rangle_r \right) \ge 0.
		\label{e:rgsol}
\ee
	Hence the BL parameterization is found to correspond to
	a Gaussian approximation to an exact solution
	of the non-relativistic hydrodynamical equations,
	with constrained values of the cooling parameters
	$\langle \Delta T/T\rangle_\tau$
	and $\langle \Delta T/T\rangle_r$.
	If the inequality in
	eq.~(\ref{e:rgsol}) is violated, the solution 
	for the density profile in eq.~(\ref{e:nbl}) will be
	still valid. In this case, 
	the radial density profile will increase first 
	before decreasing at large distances. This type of solution
	corresponds to a kind of {\it expanding ``smoke-ring"}
	in the transversal plane, or, to an {\it expanding 
	spherical shell} if the expansion is three dimensional,
	as illustrated on Figure 1 for $R_0 = 5$ fm, $m \mf^2/T_0 = 0.125$.
	For $\langle \Delta T/T\rangle_r = 0.06 < m \mf^2/T_0 $
	the density has one maximum while for 
	$\langle \Delta T/T\rangle_r = 0.2 > m \mf^2/T_0 $
	the density profile develops two maxima in any transversal direction.

	{\it Application to FOPI data} ---
	The FOPI Collaboration measured recently the proton-proton
	 correlation functions at 1.93 AGeV Ni + Ni collisions~\cite{FOPI}.
	To interpret their data, they utilized, independently,
	a version of the hydrodynamical solution, presented here.
	They assumed a linear flow profile, a Gaussian density
	distribution and a constant temperature.
	This solution corresponds to a collisionless
	Knudsen gas~\cite{cspeter}. One expects that
	a collisionless approximation breaks down for
	1.93 AGeV Ni + Ni collisions.
	Indeed, only the peak of the FOPI experimental
	proton-proton correlation functions
	was well reproduced by the collisionless model, however,
	the tails had to be excluded from the fit. From the above presented
	solution
	of the equations of the NR hydrodynamical
	systems we know,  that the temperature distribution must
	have a radial and a temporal profile
	{\it if }  a non-trivial hydro evolution is expected.
	Such a situation was also studied
	in ref.~\cite{nrt}, where it was shown that a
	temperature profile  results in a
	momentum dependence of the effective radius parameters
	of the proton-proton correlation functions and
	in a characteristic bending down of the proton-
	spectra, observable as a deviation from the exponential
	spectrum~\cite{FOPI}. Hence we predict 
	a deviation from the exponential proton spectrum and
	a momentum dependence of the effective source radii, to be
	observed in a more detailed analysis of the FOPI data.
	Note, that comparison with experimental
	data  with the help of a hydro solution	
	requires also a condition for the particle 
	freeze-out, see ref.~\cite{jnr} for
	details.	

	{\it Summary} --- 
	A new family of  simple  and
	exact solutions of the non-relativistic hydrodynamical
	equations is found for rotationally symmetric fireballs,
	for an ideal gas equation of state.
	Without radial temperature
	gradients,  the only solution of these
	equations was found to be
	a collisionless Knudsen gas lacking dynamical evolution.
	The results are generalized to expansion  
	in arbitrary number of spatial dimensions. The Zim\'anyi-Bondorf-Garpman
	solution~\cite{jnr} and the Buda-Lund hydro parameterization~\cite{3d}
	are recovered as special cases. For strong enough radial 
	decrease of the temperature,
	expanding smoke-ring type of hydro solutions are described.

	{\it Acknowledgments:}  Exciting discussions are gratefully
	acknowledged with M. Pl\"umer,
	J. Balog,  P. Csizmadia, P. L\'evai, B. Luk\'acs,
	J. Zim\'anyi and J. Zsigrai.
	This work has been supported
	by the US - Hungarian Joint Fund MAKA 652/1998,
	by the OTKA grants T026435 and T024094, the NWO - OTKA
	grant N 25487 and by a former Advanced Research Award
	of the Fulbright Foundation.


\begin{figure}
\begin{center}
\vspace*{8.cm}
\includegraphics{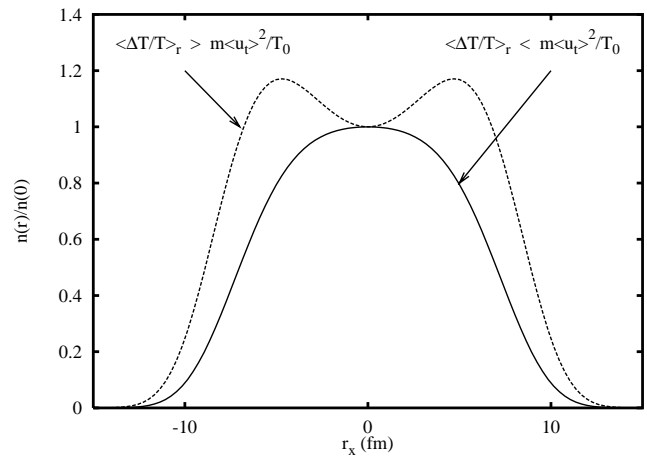}
\vspace{-1.5cm}
\end{center}
\caption{Illustration of the development of smoke-ring solutions
for large temperature gradients in 
exact solutions of non-relativistic hydrodynamics.}
\end{figure}
\end{document}